\documentclass[aps,prc,twocolumn,showpacs]{revtex4}
\usepackage{amssymb}

\usepackage{amsmath,amssymb,mathrsfs}
\usepackage{graphicx}
\usepackage{amsthm}
\usepackage{amscd}
\usepackage{bm}
\usepackage{dsfont}

\begin{document}

\title{Remarks on the origin of Castillejo-Dalitz-Dyson poles}
\author{M. I. Krivoruchenko}
\affiliation{Institute for Theoretical and Experimental Physics$\mathrm{,}$ B. Cheremushkinskaya 25 \\
117218 Moscow$\mathrm{,}$ Russia }

\begin{abstract}
Castillejo-Dalitz-Dyson (CDD) poles are known to be connected with bound states and resonances. We discuss a new 
type of  CDD pole associated with primitives i.e., poles of the $P$ matrix that correspond to zeros of 
the $D$ function on the unitary cut. Low's scattering equation is generalized for amplitudes with primitives. 
The relationship between the CDD poles and the primitives is illustrated by a description of the $S$-wave nucleon-nucleon phase shifts.
\end{abstract}


\maketitle

%%%%%%%%%%%%%%%%%%%%%%%%%%%%%%%%%%%%%%%%%%%%%%%%%%%%%%%%%%%%%%%%%

The poles introduced by Castillejo \textit{et al.} in Ref. \cite{CDD} 
are known as the Castillejo-Dalitz-Dyson (CDD) poles. They describe ambiguities in solutions to 
the Low scattering equation \cite{LOW} for amplitudes that satisfy correct 
analytical properties and unitarity. To clarify the physical meaning 
of the CDD poles, Dyson constructed a model \cite{DYSO} that demonstrates 
the relation of the CDD poles to bound states and resonances.

Some time ago, Jaffe and Low \cite{JALO} proposed a method for identifying exotic
multiquark states with primitives that appear as poles of the $P$ matrix rather 
than the $S$ matrix. The analysis, performed for scalar mesons \cite{JALO} and nucleon-nucleon
scattering \cite{JASH}, revealed primitives, in agreement with expectations from 
the MIT bag model. A dynamical model of the $P$ matrix
was developed by Simonov \cite{SIMO} and was applied to the description of
nucleon-nucleon scattering \cite{SIMO,SIMO84,BHGU,FALE,Benjamins:1988be,NARO94}.
Other microscopic models of nucleon-nucleon forces have also been discussed \cite{MULD,BENJA,AFKS}.

The recent interest in the problem of nucleon-nucleon interactions is connected to new 
constraints on the equation of state (EOS) of nuclear matter, obtained from collective 
flow data and subthreshold kaon production in heavy-ion collisions \cite{DANI02,FUCH07} 
as well as astrophysical observations of massive neutron stars \cite{BARR05,OZEL06}.
One-boson exchange models are in reasonable agreement 
with the laboratory data but predict surprisingly low masses for
neutron stars in the $\beta$ equilibrium \cite{ISHI08,SCHA08}. 
Microscopic models can provide better insight into
the short-range dynamics of nucleon-nucleon interactions and 
the high-density EOS.

In this Brief Report, we clarify the link between the CDD poles and the primitives,
which can be useful for modeling the nucleon-nucleon interactions 
in the $P$-matrix formalism.

CDD poles arise when the interaction of particles includes intermediate states that are internally
different from combined-particle states. These are discrete eigenstates of the system and basically
form other channels in the scattering problem. They have also been referred to as elementary
particle or compound states.

%%%%%%%%%%%%%%%%%%%%%%%%%%%%%%%%%%%%%%%%%%%%%%%%%%%%%%%%%%%%%%%%%

In the Dyson model, one starts from the scattering of two particles, e.g., 
a nucleon and a pion. A nucleon can absorb a pion and can turn into an excited compound
state $N_{\alpha }$ of mass $M_{\alpha }>\sqrt{s_{0}} = m + \mu,$ where $m$
and $\mu$ are the nucleon and the pion masses. The $D$ function of the process
can be written as
\begin{equation}
D(s)=\Lambda (s)-\Pi (s),  \label{DFUN}
\end{equation}
where, in the relativistic notations, 
\begin{eqnarray}
\Lambda ^{-1}(s) &=&\sum_{\alpha}\frac{ g_{\alpha }^{2}}{s-M_{\alpha }^{2}},
\label{LAMB} \\
\Pi (s) &=&-\frac{1}{\pi }\int_{s_{0}}^{+\infty }\Phi _{2}(s^{\prime }) 
\frac{\mathcal{F}^{2}(s^{\prime })}{s^{\prime }-s}ds^{\prime }.  \label{POLA}
\end{eqnarray}
Here, $\Phi _{2}(s) = \pi k/\sqrt{s}$ is the relativistic two-body phase space,
$k$ is the center-of-mass momentum, $g_{\alpha }$ is the coupling constant, 
and $\mathcal{F}(s)$ is the form factor of the $N_{\alpha }N\pi$ vertex.
The $S$ matrix has the form 
\begin{equation}
S=e^{2i\delta (s)}=\frac{D(s-i0)}{D(s+i0)}.  \label{SMAT}
\end{equation}
The poles of $\Lambda (s)$ are the CDD poles. They are located between
the zeros of $\Lambda (s)$ (i.e., between $M_{\alpha }^{2}$ and $M_{\alpha + 1}^{2}$).

The $D$ function constructed in such a way is the generalized $R$ function 
\cite{CDD}. It has no complex zeros on the first Riemann sheet of the complex 
$s$ plane. It also has no zeros on the real half axis $(-\infty ,s_{0})$, which
corresponds to bound states, provided $D(s_{0})<0$ and $s_{0} < M_{\alpha }^{2}$.

The simple roots of the equation 
\begin{equation}
D(s)=0,  \label{DNULL}
\end{equation}
located on the second Riemann sheet below the unitary cut, are identified as
resonances. In the limit of small $g_{\alpha }$, roots of Eq.~(\ref{DNULL}) 
are localized in the neighborhood of $s = M_{\alpha }^{2}$. $\Re s$ gives the
renormalized resonance mass, while $\Im s$ determines the decay
width $\Gamma_{\alpha} = g_{\alpha}^2 \Im D(M_{\alpha }^2)/M_{\alpha }$.

At the CDD poles $\delta (s)=0 \mod(\pi)$, the slope of the phase is
positive. If $s_{\gamma}$ is a CDD pole, then Eq.~(\ref{LAMB})
gives $\Lambda^{-1}(s_{\gamma}) = 0$ and $\Lambda^{-1}(s_{\gamma})^{\prime}
< 0$. By expanding the $D$ function around $s = s_{\gamma}$ and by using 
Eq.~(\ref{SMAT}), one finds 
\[
\delta(s_{\gamma})^{\prime} = - \Im D(s_{\gamma})
\Lambda^{-1}(s_{\gamma})^{\prime} > 0. 
\]
Such behavior is in agreement with the Breit-Wigner formula according to which
isolated resonances drive the phase shift up. In potential scattering, 
an increasing phase is associated with attraction.

The Dyson model therefore applies to systems with attraction
where scattering phase shifts increase with increasing energy. 

%%%%%%%%%%%%%%%%%%%%%%%%%%%%%%%%%%%%%%%%%%%%%%%%%%%%%%%%%%%%%%%%%

The nucleon-nucleon phase shifts, conversely, decrease with increasing 
energy and provide evidence for repulsion.

In Refs.~\cite{CDD,LOW,DYSO} $\Im D(s)$ is strictly positive.
Softening this constraint to $\Im D(s) \geq 0$ allows extension of 
the Dyson model to systems with repulsion:

Let us consider the scattering of two nucleons through compound states, dibaryons, 
with form factors $\mathcal{F}(s)$ 
that have a simple zero at $s = s_{p} > s_{0} = 4m^{2}$. 
Consequently, $\Im D(s)\sim (s-s_{p})^{2}$. Such behavior is
presupposed in the quark compound bag (QCB) model developed by Simonov for
the description of nucleon-nucleon interactions \cite{SIMO}. 
In the QCB model, the separable potential generated by the compound 6-quark bags 
is restricted to the bag surfaces. The $S$-wave form factor has the form 
$\mathcal{F}(s)=\sin(kb)/kb$, where $k$ is the center-of-mass momentum 
and $b$ is the effective interaction radius. In relativistic notation, 
the $D$ function of the model has the form of Eq.~(\ref{DFUN}), 
while $\Lambda (s)$ and the self-energy operator $\Pi (s)$ are equivalent 
to Eqs. (\ref{LAMB}) and (\ref{POLA}), respectively. 

This analogy allows the techniques developed in Ref.~\cite{CDD} 
to be used to parametrize nucleon-nucleon scattering amplitudes
with functions that have the correct analytical properties. 

If $\Re D(s_{p}) \neq 0$, the phase touches, at $s=s_{p}$, one of the 
$\delta (s)=0 \mod(\pi )$ levels without crossing. However, if Eq.~(\ref{DNULL})
holds at $s=s_{p}$ for both the real and the imaginary parts, the phase crosses one
of the levels $\delta(s)=0 \mod(\pi)$ \textit{with a negative slope}. This can
be verified by expanding $D(s)$ around $s = s_{p}$. By taking Eq.~(\ref
{SMAT}) and the conditions $\Im D(s_{p})^{\prime \prime } > 0$ and $\Re
D(s_{p})^{\prime } >0$ into account, one gets 
\[
\delta (s_{p})^{\prime }=-\frac{\Im D(s_{p})^{\prime \prime }}{2\Re
D(s_{p})^{\prime }} < 0. 
\]
In potential scattering, a negative slope of the phase shift is
associated with repulsion.

%%%%%%%%%%%%%%%%%%%%%%%%%%%%%%%%%%%%%%%%%%%%%%%%%%%%%%%%%%%%%%%%%

The Low scattering equation \cite{LOW} is modified in the presence of primitives. 
In the systems with $\Pi(s)$ given by Eq.~(\ref{POLA}),
the scattering amplitude $A(s)=e^{i\delta (s)}\sin \delta (s)$ can be represented as follows: 
\begin{equation}
A(s)= - \frac{\Phi _{2}(s)\mathcal{F}^{2}(s)}{D(s)}.
\end{equation}
This amplitude obeys the generalized Low scattering equation
\begin{widetext}
\begin{equation}
\frac{A(s)}{\Phi _{2}(s)\mathcal{F}^{2}(s)} = \frac{1}{\pi }\int_{s_{0}}^{+\infty }
\frac{|A(s^{\prime })|^{2}}{\Phi _{2}(s^{\prime })\mathcal{F}^{2}(s^{\prime })}
\frac{ds^{\prime }}{s^{\prime }-s} - \sum_{b}\frac{C_{b}}{s-s_{b}} - \sum_{p}\frac{C_{p}}{s-s_{p}} - C,
\label{GLOW}
\end{equation}
\end{widetext}
which is essentially the dispersion integral representation for the inverse 
denominator function $D(s)$ that accounts for the poles, which correspond 
to the bound states and primitives. 
$A(s)$ and $\mathcal{F}(s)$ have simple zeros at $s=s_{p}$, so the
integrand in Eq.~(\ref{GLOW}) is a regular function at $s^{\prime }=s_{p}$.
The bound states and the primitives generate poles at $s_{b}<s_{0}$ and $s_{p}>s_{0}$ 
on the real axis, the coefficient
\[
C_{p}=-\frac{2A(s_{p})^{\prime }}{\Phi _{2}(s_{p})\mathcal{F}%
^{2}(s_{p})^{\prime \prime }}
\]
is positive. 

%%%%%%%%%%%%%%%%%%%%%%%%%%%%%%%%%%%%%%%%%%%%%%%%%%%%%%%%%%%%%%%%%

In the QCB model, the $P$ matrix takes the form
\begin{equation}
P = P_{free} + \kappa ^{-1}\Lambda ^{-1},
\end{equation}
where $P_{free}$ is the free $P$ matrix. 
For the $S$ wave, $P_{free} = kb\cot(kb)$ and $P = kb\cot(kb + \delta(s))$. 
The value of $\kappa$ is fixed by the normalization of $D(s)$.
The compound states of masses $M_{\alpha }$ show up as poles
of the $P$ matrix. The poles of the $P$ matrix split into 
two groups according to their physical nature:

The first group is related to the bound states and resonances.

One bound state always exists at $D(s_{0})>0$. Additional bound states 
can be generated by compound states with masses $M_{\alpha }<\sqrt{s_{0}}$.

A characteristic feature of a resonance is the condition $\mathcal{F}(s)
\neq 0$ in the neighborhood of $s=M_{\alpha }^{2}$. Equation (\ref{DNULL}) can
then be used to find a simple pole of the $S$ matrix. 
The roots of Eq.~(\ref{DNULL}) that lie on the real half axis 
$(-\infty ,s_{0})$ of the second Riemann sheet are virtual states that
can be related to the compound states also.

The poles of the second group are related to roots of Eq.~(\ref{DNULL}) 
\textit{on the unitary cut} in the neighborhood of $s=M_{\alpha }^{2}$. 
Such poles do not
show up as $S$-matrix poles and cannot be treated as resonances. 
They are called primitives according to Jaffe and Low \cite{JALO}. 
If a resonance moves from the second Riemann sheet to the unitary cut, 
its singular effect on the $S$ matrix cancels out. 
As distinct from resonances, primitives drive the
phase shift down and mimic repulsion.

\begin{figure} [!htb]
\includegraphics[angle = 0,width=0.250\textwidth]{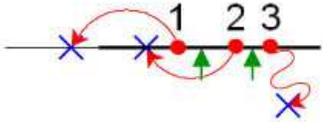}
\caption{
(Color online)
$D$-function zeros in the complex $s$ plane. The unitary cut is shown by 
a bold solid line. Compound states 1, 2, and 3 are eigenstates of the
free Hamiltonian. Upon switching on the $s$ channel interaction with the continuum, zeros move to new positions 
shown by crosses. Compound states 1, 2, and 3 become bound state, primitive, 
and resonance, respectively. 
A pair of the CDD poles that squeezes compound state 2 of the primitive type 
is shown by arrows.
}
\label{fig:compound}
\end{figure}

In the Dyson model, there exist at most one bound state and at most one resonance, 
which are not associated with the CDD poles. In the QCB model, 
there are CDD poles related to primitives that do not give rise to bound states or resonances. 
The neighboring CDD poles squeeze masses of compound states that become 
bound states, resonances, or primitives when coupling to the continuum is switched on. 
This is illustrated in Fig. \ref{fig:compound}. 

%%%%%%%%%%%%%%%%%%%%%%%%%%%%%%%%%%%%%%%%%%%%%%%%%%%%%%%%%%%%%%%%%%%%%%%%%%%%
%%%%%%%%%%%%%%%%%%%%%%%%%%%%%%%%%%%%%%%%%%%%%%%%%%%%%%%%%%%%%%%%%%%%%%%%%%%%
%%%%%%%%%%%%%%%%%%%%%%%%%%%%%%%%%%%%%%%%%%%%%%%%%%%%%%%%%%%%%%%%%%%%%%%%%%%%

%%%%%%%%%%%%%%%%%%%%%%%%%%%%%%%%%%%%%%%%%%%%%%%%%%%%%%%%%%%%%%%%%%%%%%%%%%%%
%%%%%%%%%%%%%%%%%%%%%%%%%%%%%%%%%%%%%%%%%%%%%%%%%%%%%%%%%%%%%%%%%%%%%%%%%%%%
%%%%%%%%%%%%%%%%%%%%%%%%%%%%%%%%%%%%%%%%%%%%%%%%%%%%%%%%%%%%%%%%%%%%%%%%%%%%

%%%%%%%%%%%%%%%%%%%%%%%%%%%%%%%%%%%%%%%%%%%%%%%%%%%%%%%%%%%%%%%%%

The $S$-wave nucleon-nucleon scattering can be considered as an
example of dynamics influenced by the CDD poles that are connected 
to primitives. 

The model we discuss is the relativistic extension of the QCB model. The $P$%
-matrix formalism is recovered with
\begin{equation}
\mathcal{F}(s)=\left( \frac{s}{s_{0}}\right) ^{1/4}\frac{\sin (kb)}{kb}.
\end{equation}
Equation (\ref{POLA}) for $\kappa = 2 m b/\pi$ gives 
\begin{equation}
\kappa \Pi (s+i0) = -\frac{\sin (kb)}{kb}e^{ikb}.
\end{equation}

In addition to the interaction through compound states, we introduce a
contact interaction. This amounts to a redefinition of $\Lambda^{-1}(s)$ as
compared to Eq.~(\ref{LAMB}). In the case of one CDD pole, the most general
expression for $\Lambda^{-1}(s)$ becomes
\begin{equation}
\kappa ^{-1} \Lambda ^{-1}(s) = c_{p}(\frac{r_{p}}{s-s_{p}}-\frac{r_{p}}{%
s_{d}-s_{p}})-\frac{1}{\gamma },  \label{INVLAMB}
\end{equation}
where $c_{p}$ is a free parameter such that $\kappa c_{p}r_{p} = g_{\alpha}^{2}$, 
$r_{p} = 8(\pi/b)^{2}$ is the residue of $P_{free}$ in the $S$ wave, and 
$s_d = M^2_{d}$ is the deuteron pole or the threshold. The $D$ function with the contact
interaction remains the generalized $R$ function.

%%%%%%%%%%%%%%%%%%%%%%%%%%%%%%%%%%%%%%%%%%%%%%%%%%%%%%%%%%%%%%%%%

In the $^{3}S_{1}$ channel, the phase shift vanishes at 
$T_{lab}=354$ MeV. The $S$ matrix according to Eq.~(\ref{SMAT}) 
is unit in two cases: $\Lambda (s)=\infty $ and $\Im D(s)\equiv -\Im \Pi (s)=0.$ The poles of 
$\Lambda(s) $ are the CDD poles. At the CDD poles, the slope of the phase is
positive, which corresponds to attraction. The second case $\Im \Pi (s)=0$ gives repulsion.

$T_{lab}=354$ MeV is equivalent to $k=408$ MeV. The equation $\mathcal{F}(s)=0$ gives 
$kb=\pi $. Thus, we determine $b=1.52$ fm. Since $\mathcal{F}(s_{p})=0$ if and
only if $\Re \Pi (s_{p})=0$, Eq.~(\ref{DNULL}) simplifies to $\Lambda
(s_{p})=0$. $\Lambda (s)$ vanishes when $s = M_{\alpha }^{2}$. 
The compound state shows up as the primitive of mass $M_{\alpha }=2\sqrt{k^{2}+m^{2}}=2047$ MeV.

The parametrization ensures the existence of the deuteron pole for 
\begin{equation}
\gamma = - \kappa \Pi (s_{d}) > 0.
\end{equation}

Unphysical zeros of the $D$ function are eliminated by constraining $c_{p}$. 
One can easily show that $\Im D(s) \sim \Im s$,
and that the coefficient of proportionality is positive for positive $c_{p}$. 
In this case, $D(s)$ has no zeros for $\Im s \neq 0$. 
The real half axis $(-\infty, s_0)$ remains to be checked. 
The derivative $D(s)^{\prime }$ is positive
below the threshold. $D(s)$ crosses the real axis at $s=s_{d}<s_0$. 
This is the unique zero of the $D$ function, provided $\Lambda (s)$ has no poles 
for $s < s_0$. Let us investigate the zeros of $\Lambda ^{-1}(s)$. Since $%
\kappa ^{-1} \Lambda^{-1}(s_{d})=-1/\gamma <0$, $\Lambda ^{-1}(s)^{\prime
}$ $<0$ and $\Lambda^{-1}(s)$ has no poles for $s<s_{p}$ by construction,
the condition $\Lambda ^{-1}(-\infty )<0$ is sufficient to exclude
unphysical zeros. Finally, $c_{p}$ satisfies the constraint 
\begin{equation}
0 < c_{p} < c_{p}^{\max }=\frac{s_{p}-s_{d}}{\gamma r_{p}}.  \label{CONS}
\end{equation}

In Fig.~\ref{fig:3S11S0}~(a), we show the $^{3}S_{1}$ phase shift versus the
proton kinetic energy for $c_{p} = 0.9c_{p}^{\max }$. This is 
compared to the partial wave analysis data provided by Ref.~\cite{PHAS}. 
For the $pn$ system, $s = s_{0} + 2m_{n}T_{lab}$, where $m_{n}$ is neutron mass. The
CDD pole is located at $M = 3203$ MeV. 
The pion production threshold is at $T_{lab}=280$ MeV and the
inelasticity is small up to $\sim 350$ MeV.

In Fig.~\ref{fig:3S11S0}~(c), we show $\Re D(s)$ and $\Im D(s)$ as functions
of $T_{lab}$. $\Re D(s)$ has one zero below $s_{0}$, which corresponds to
the deuteron. The second zero at $s>s_{0}$ with $\Im D(s) \neq 0$ and a negative slope 
of $\Re D(s)$ 
ensures the crossing of the level $\delta(s) = \pi$. The third zero corresponds to the
primitive.

%%%%%%%%%%%%%%%%%%%%%%%%%%%%%%%%%%%%%%%%%%%%%%%%%%%%%%%%%%%%%%%%%

In the $^{1}S_{0}$ channel, the phase shift vanishes at $T_{lab}=265$ MeV. 
The same arguments as before give $b=1.76$ fm and $M_{\alpha} = 2006$ MeV.

$\Lambda^{-1}(s)$ has the form of Eq.~(\ref{INVLAMB}), with $s_{d}$ replaced
by $s_{0}=4m^{2}.$ Near the threshold, $\kappa D(s) = -\gamma + 1 + ikb +
\ldots$ From the other side, $D(s)\sim 1-i\delta (k)+...=1-ika+...,$ where $%
a=23.56$ fm is the scattering length. One has to require 
\begin{equation}
\gamma = 1 + \frac{b}{a}.
\end{equation}

$D(s)$ has no zeros for complex values of $s$. Its derivative is
positive for real $s < s_{0}$. To avoid unphysical zeros, 
it is sufficient to require 
\[
\kappa D(-\infty )=\frac{1}{-\frac{c_{p}r_{p}}{s_{0}-s_{p}}-\frac{1}{%
\gamma }}< \kappa D(s_{0})=1-\gamma <0. 
\]
The second inequality is fulfilled, and the first one gives 
$c_{p}<\min (c_{p}^{\max },{c_{p}^{\max }}/{(\gamma -1)})$.
Since $b \ll a$, this reduces to Eq.~(\ref{CONS}) with $s_d$ replaced by $s_0$.

In Fig.~\ref{fig:3S11S0}~(b), we show our fit of
the $pn$ $^{1}S_{0}$ phase shift with $c_{p}=0.9c_{p}^{\max }$
compared to the experimental data \cite{PHAS}. 
The CDD pole occurs at $M = 2916$ MeV.

Shown in Fig.~\ref{fig:3S11S0}~(d) are the real and the imaginary parts of the $D$ function
versus the proton kinetic energy. 

In Figs. \ref{fig:3S11S0}~(c,d), the real and the imaginary parts of the $D$ functions 
vanish at $s=s_{p}$. These are signatures of the primitives, 
along with crossing the levels $\delta(s) = 0$ with negative slopes on Figs. \ref{fig:3S11S0}~(a,b).
 
The values of $b$ and $M_{\alpha}$ are close to those obtained in Refs. \cite{JASH,SIMO}.

Benjamins and van Dijk \cite{BENJA} used the hybrid Lee model with one compound state 
in each channel to describe the nucleon-nucleon $S$-wave phase shifts below $T_{lab} = 500$ MeV 
and to reproduce parameters related to the deuteron and the virtual $^1 S_{0}$ state.
The model does not have explicit CDD poles and primitives. However, it can be reformulated 
in terms of the QCB model with one CDD pole and two compound states that correspond
to the primitive and a high-mass resonance \cite{MIK10}.

%%%%%%%%%%%%%%%%%%%%%%%%%%%%%%%%%%%%%%%%%%%%%%%%%%%%%%%%%%%%%%%%%
%%%%%%%%%%%%%%%%%%%%%%%%%%%%%%%%%%%%%%%%%%%%%%%%%%%%%%%%%%%%%%%%%
%%%%%%%%%%%%%%%%%%%%%%%%%%%%%%%%%%%%%%%%%%%%%%%%%%%%%%%%%%%%%%%%%
\onecolumngrid
\begin{widetext}
\vspace{-9mm}
\begin{figure} [!htb] %[!htb]
  \centering
  \includegraphics[angle =   0,width=0.382\textwidth]{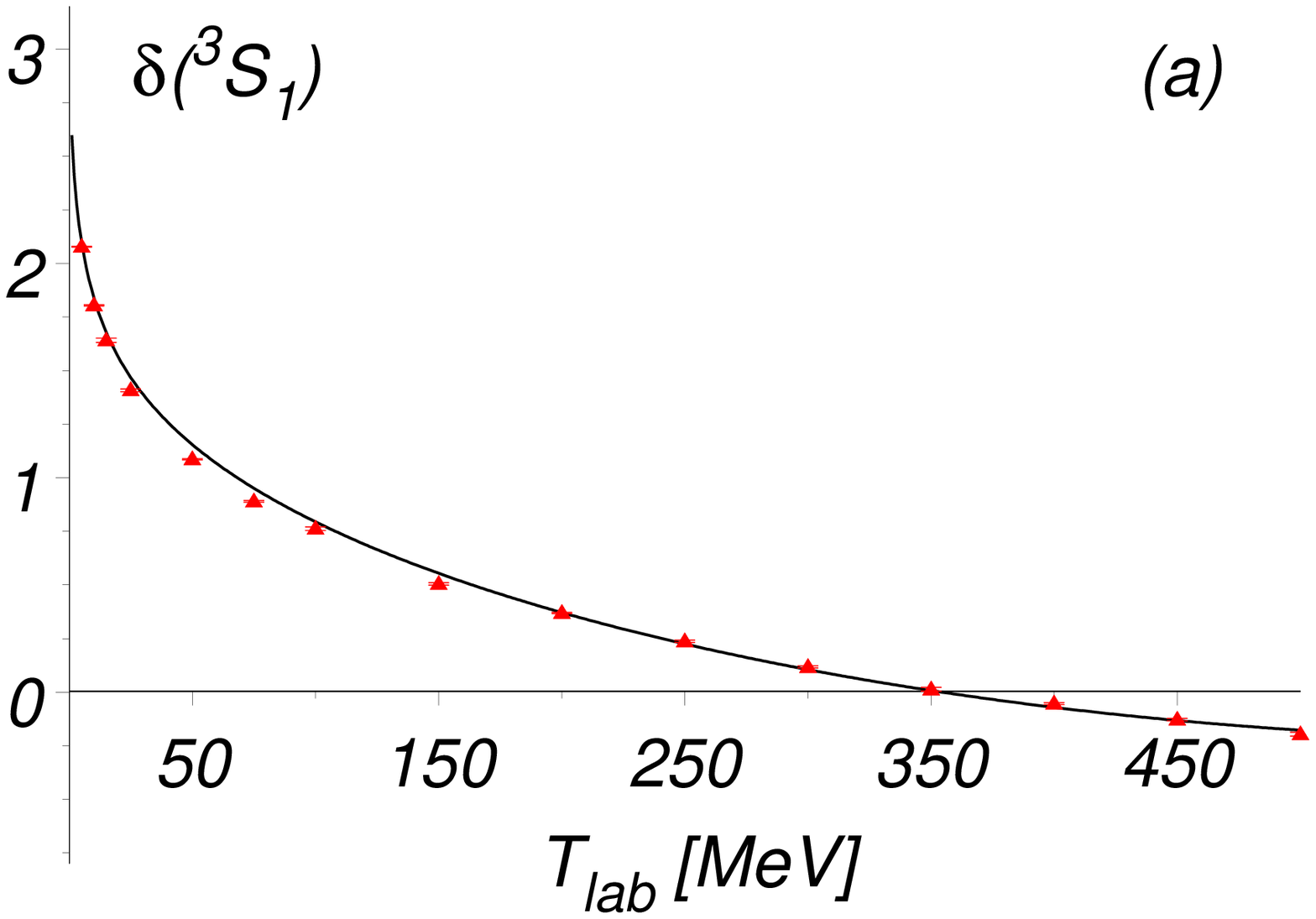}~~~~~~~~~~~~~~~~~~~~~
  \includegraphics[angle =   0,width=0.382\textwidth]{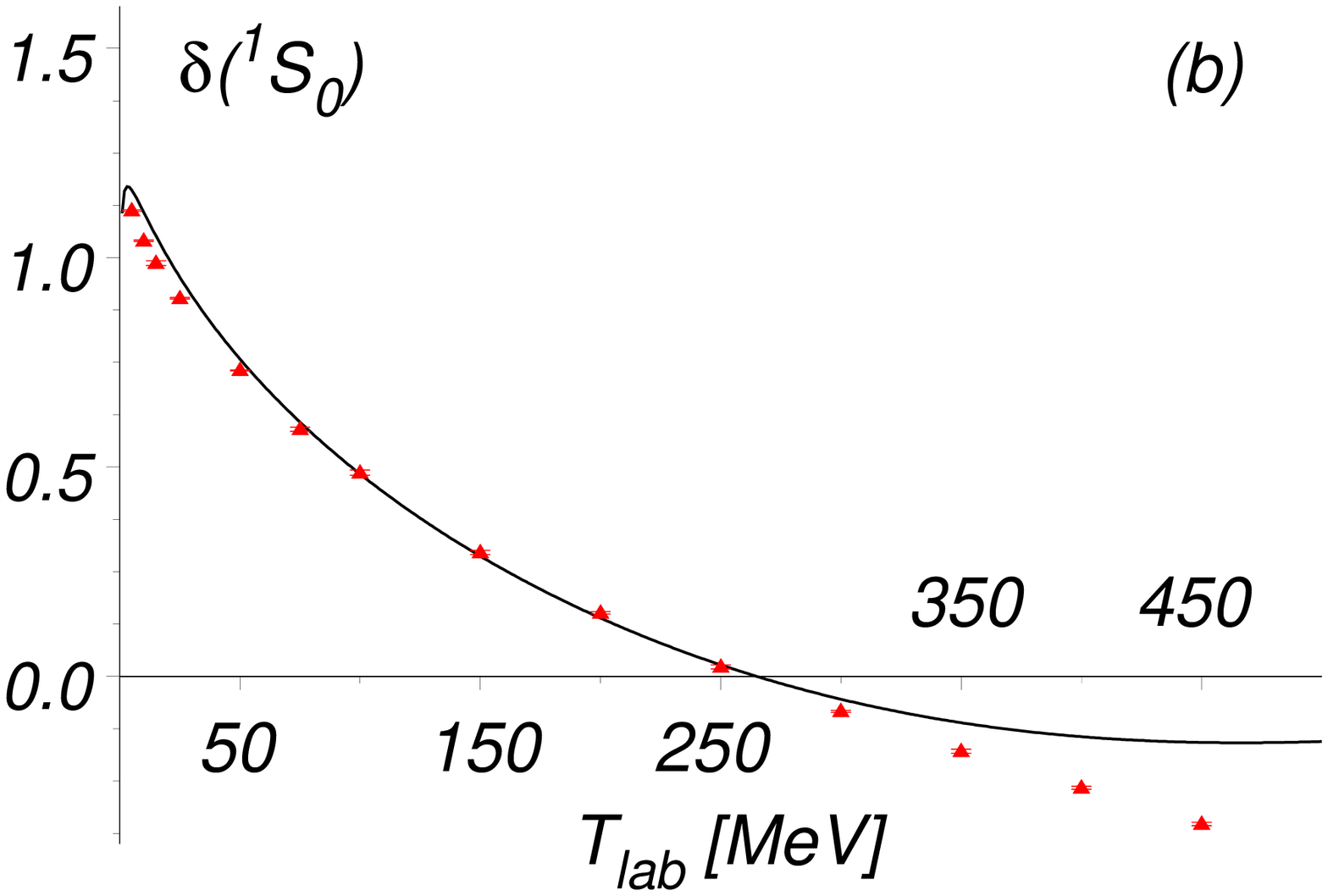} \\
  \includegraphics[angle =   0,width=0.413\textwidth]{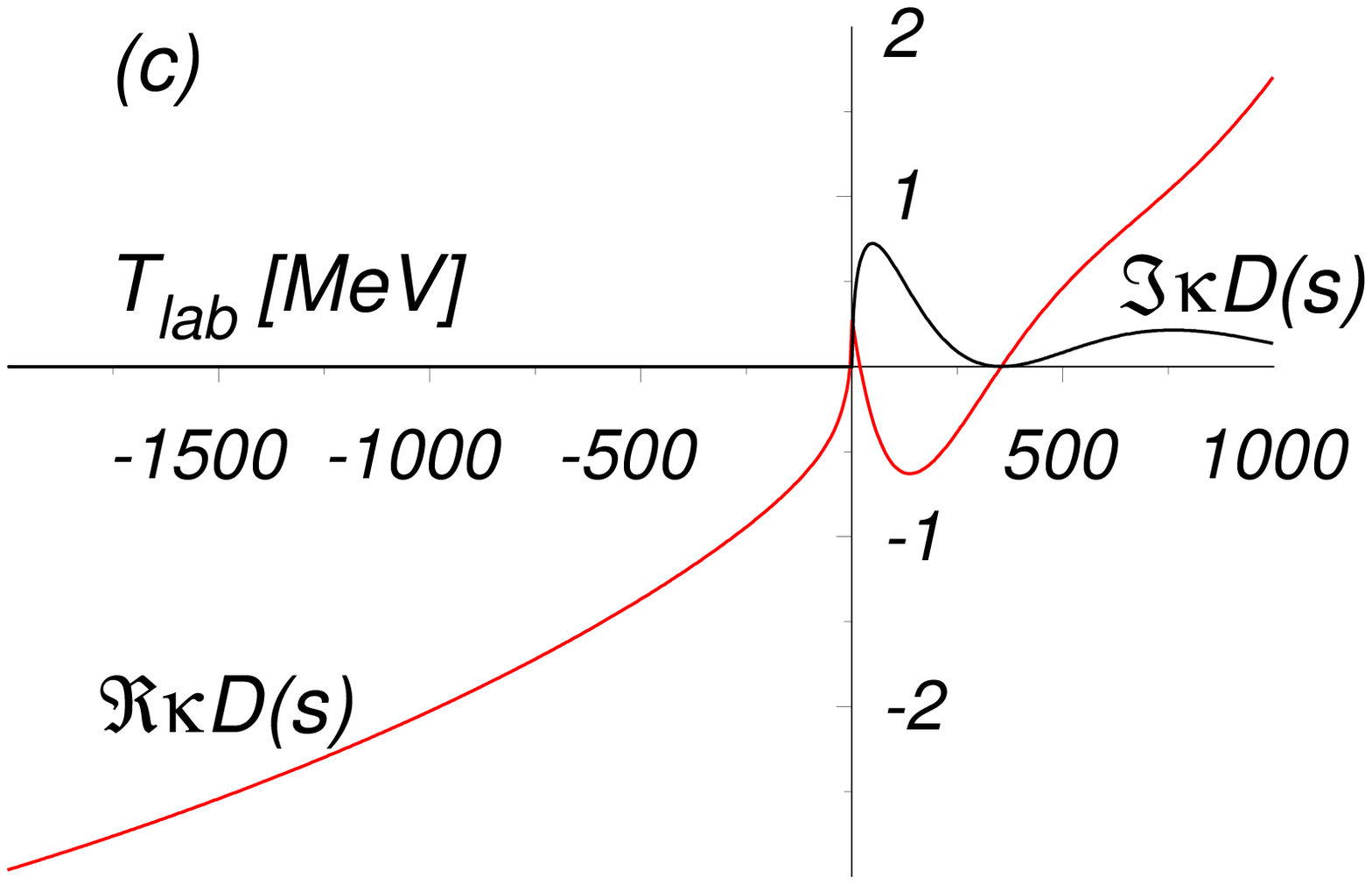}~~~~~~~~~~~~~~~~
  \includegraphics[angle =   0,width=0.413\textwidth]{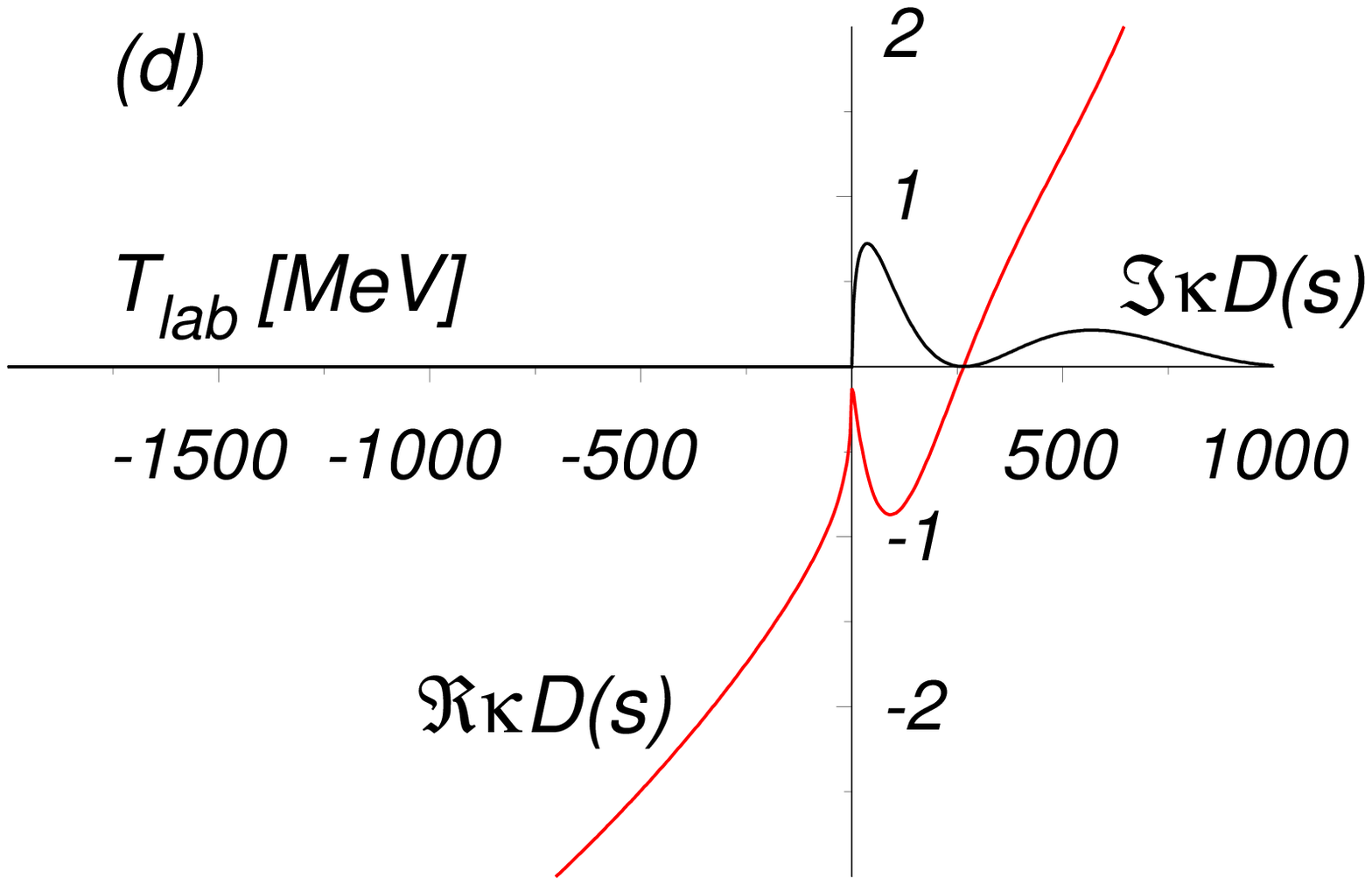}~~~~
\vspace{-9mm}
\caption{ (Color online)
$^{3}S_{1}$ and $^{1}S_{0}$ scattering phase shifts in radians (upper panel) and real and imaginary parts of 
the $D$ functions (lower panel) 
versus the proton kinetic energy. The solid curves are parametrizations 
within the relativistic QCB model. 
The experimental phase shifts \cite{PHAS} are shown by triangles. 
}
\label{fig:3S11S0}
\end{figure}
\end{widetext}
\twocolumngrid
%%%%%%%%%%%%%%%%%%%%%%%%%%%%%%%%%%%%%%%%%%%%%%%%%%%%%%%%%%%%%%%%%
%%%%%%%%%%%%%%%%%%%%%%%%%%%%%%%%%%%%%%%%%%%%%%%%%%%%%%%%%%%%%%%%%
%%%%%%%%%%%%%%%%%%%%%%%%%%%%%%%%%%%%%%%%%%%%%%%%%%%%%%%%%%%%%%%%%

%%%%%%%%%%%%%%%%%%%%%%%%%%%%%%%%%%%%%%%%%%%%%%%%%%%%%%%%%%%%%%%%%

Resonances and primitives do not exist as asymptotic states. In Feynman
diagrams, propagators of primitives $1/(s-M_{\alpha }^{2})$ are multiplied
by form factors $\mathcal{F}(s)$. Such combinations do not have poles at 
$s=M_{\alpha }^{2}$. Primitives thus do not propagate, though they influence the dynamics.

Summarizing, the physical meaning of the CDD poles was revisited. 
In the general case, the neighboring CDD poles squeeze 
masses of compound states related to bound states, %virtual states,
resonances, or primitives. The primitives are $P$-matrix poles 
associated with zeros of the $D$ function on the unitary cut, which do not show up as poles of the $S$ matrix. 
The Low scattering equation was generalized for amplitudes with primitives.
The primitive-type CDD poles occur in systems with repulsion.
In the $^{3}S_{1}$ and $^{1}S_{0}$ nucleon-nucleon channels, 
the CDD poles at $M = 3203$ MeV and $M = 2916$ MeV are associated with
the primitives at $M_{\alpha } = 2047$ MeV and $M_{\alpha} = 2006$ MeV, respectively.
The model we used ensures that the $D$ function
has the correct analytical properties on the first Riemann sheet of the complex $s$ plane 
and provides the partial wave amplitudes
that satisfy the generalized Low scattering equation.

The author is grateful to Yu.~A.~Simonov for helpful discussions and 
I.~M.~Narodetsky for reading the manuscript and for useful remarks.
This work is supported by grant of Scientific Schools 
of Russian Federation No. 4568.2008.2,
RFBR grant No. 09-02-91341, and DFG grant No. 436 RUS 113/721/0-3.

%%%%%%%%%%%%%%%%%%%%%%%%%%%%%%%%%%%%%%%%%%%%%%%%%%%%%%%%%%%%%%%%%

\end{document}